\documentstyle{article}
\begin{document}

\begin{center}
{\Large Two-dimensional solitary pulses in driven diffractive-diffusive
complex Ginzburg-Landau equations}

\bigskip

Hidetsugu Sakaguchi

Department of Applied Science for Electronics and Materials,
Interdisciplinary Graduate School of Engineering Sciences, Kyushu
University, Kasuga 816-8580, Japan

\smallskip

Boris A. Malomed\\[0pt]

Department of Interdisciplinary Studies, Faculty of Engineering, Tel Aviv
University, Tel Aviv 69978, Israel

\bigskip

Abstract
\end{center}

Two models of driven optical cavities, based on two-dimensional
Ginzburg-Landau equations, are introduced. The models include loss, the Kerr
nonlinearity, diffraction in one transverse direction, and a combination of
diffusion and dispersion in the other one (which is, actually, a temporal
direction). Each model is driven either parametrically or directly by an
external field. By means of direct simulations, stable completely localized
pulses are found (in the directly driven model, they are built on top of a
nonzero flat background). These solitary pulses correspond to
spatio-temporal solitons (``light bullets'') in the optical cavities. Basic
results are presented in a compact form as stability regions in a full
three-dimensional parameter space of either model. The stability region is
bounded by two surfaces; beyond the left one, any two-dimensional (2D) pulse
decays to zero, while quasi-1D\ pulses, representing spatial solitons in the
optical cavity, are found beyond the right boundary. The spatial solitons
are found to be stable both inside the stability region of the 2D pulses
(hence, bistability takes place in this region in the two models) and beyond
the right boundary of this region (although they are not stable everywhere).
Unlike the spatial solitons, their quasi-1D counterparts in the form of
purely temporal solitons are always subject to modulational instability,
which splits them into an array of 2D pulses, that further coalesce into two
final pulses. A uniform nonzero state in the parametrically driven model is
also modulationally unstable, which leads to formation of many 2D pulses
that subsequently merge into few ones.

\section{Introduction}

Complex Ginzburg-Landau (CGL) equations is a class of universal models
describing pattern formation in media of various physical nature, combining
nonlinearity with dissipative and dispersive linear properties. They find
important applications to traveling-wave convection (see Ref. \cite{Kurths}
and references therein), fiber optics \cite{Agrawal}, and in other fields.
Nonlinear terms in a CGL equation, as well the linear ones, may be both
dissipative, accounting for loss and gain, and conservative.

An important class of patterns that appear in different versions of the CGL
equations are solitary pulses (SPs). In particular, a GL model with the
cubic-quintic (CQ) nonlinearity was first introduced by Petviashvili and
Sergeev \cite{PetSer} in the two-dimensional (2D) case, with an objective to
construct stable fully localized 2D pulses. The nonlinearity of the CQ type
was necessary, as the stability of a pulse requires, first of all, the zero
background to be stable, which implies that the linear part of the equation
could not contain gain, hence a gain term had to be placed in the cubic
nonlinearity. Finally, to provide for the overall stability of the model,
the cubic gain had to be capped with a quintic lossy term.

Stable SPs (``solitons'') in 2D CGL equations of various types were found in
Refs. \cite{Deissler}. Stable axisymmetric solitons with an internal
vorticity (``spin'') have been found too in the isotropic CQ CGL equation 
\cite{Bucharest}. On the other hand, the 2D CGL equation may find its most
plausible experimental realization as a model governing evolution of optical
spatiotemporal pulses in a planar nonlinear waveguide (2D optical cavity);
in that case, however, the equation is strongly anisotropic due to the
natural difference between the propagation direction ($z$) and the
transverse one ($x$) along which the electromagnetic wave diffracts. In a
rather general form, the corresponding anisotropic 2D model was put forward
in Ref. \cite{we}: 
\begin{equation}
iu_{z}+\frac{1}{2}u_{xx}-(\frac{i}{2}+\beta )u_{tt}+\left[ iu+\left(
1-i\gamma _{1}\right) |u|^{2}u+i\gamma _{2}|u|^{4}u\right] =0.  \label{CQ}
\end{equation}
In this equation, $u(z,x,t)$ is the local amplitude of the electromagnetic
field, $t\equiv T-z/V_{0}$ is the so-called reduced time, where $T$ is time
proper, and $V_{0}$ is the group velocity of the carrier wave propagating in
the $z$--direction \cite{Agrawal}. Further, coefficients in front of the
terms accounting for the transverse diffraction ($\frac{1}{2}u_{xx}$),
linear loss ($iu$), optical filtering ($-\frac{1}{2}iu_{tt}$, which formally
looks like diffusion in the $t$--direction), and Kerr nonlinearity ($
|u|^{2}u $) are all normalized to be $1$. The remaining free parameters $
\beta $, $\gamma _{1}$, and $\gamma _{2}$ control the temporal dispersion,
nonlinear cubic gain, and quintic loss ($\gamma _{1}$ and $\gamma _{2}$ must
be positive, while $\beta $ may have either sign or be equal to zero). SP
solutions to Eq. (\ref{CQ}) in the form $u(z,x,t)=\exp (ikz)$\thinspace $
v(x,t)$, with the function $v(x,t)$ exponentially vanishing at $\left|
x\right| \rightarrow \infty $ and $\left| t\right| \rightarrow \infty $,
were found in Ref. \cite{we}, and their stability region was identified in
the corresponding parameter space.

The model based on Eq. (\ref{CQ}) assumes that the planar waveguide carrying
the field is equipped with uniformly distributed linear {\it 
bandwidth-limited} gain (which induces the filtering term) combined with a
saturable absorber \cite{Agrawal}, that together give rise to the CQ
nonlinearity. For applications to nonlinear optics, it is also quite
interesting to consider models where wave patterns are supported, instead of
the intrinsic gain, by an external pump field, that leads to {\em driven}
CGL equations, in which it is sufficient to take into regard the ordinary
cubic (Kerr) nonlinearity, while the linear diffraction, dispersion, and
filtering terms remain the same as in Eq. (\ref{CQ}). In fact, two different
driving terms may appear in the accordingly modified equation: a parametric
drive and direct forcing, see below. It is quite interesting to find 2D
spatiotemporal SPs in the driven optical models, as they have a real chance
to be observed in the experiment, and are also of interest in their own
right as completely localized pulses in the anisotropic 2D\ CGL equations of
the new type.

In this work, we find these pulses and demonstrate their stability. Besides
that, we also study quasi-1D solitons of two different types, which
represent spatial and temporal solitons in the underlying optical models. It
will be demonstrated that the spatial solitons are stable in a broad
parametric region (although not everywhere); in particular, they coexist as
stable attractors with the stable 2D pulses, hence both models are bistable
in the corresponding regions. On the contrary to that, the temporal solitons
are always modulationally unstable, and simulations show that the
instability splits them into an array of 2D pulses, which subsequently
coalesce into two 2D solitons. A uniform nonzero state in the parametrically
driven model is also subject to modulational instability, which leads to a
formation of a large number of 2D SPs; later, they merge into few final 2D
solitons.

The rest of the paper is organized as follows. The two versions of the
driven model are introduced in section 2, and SP solutions found in them are
displayed in section 3. The results of extensive numerical simulations are
summarized in the form of explicit stability diagrams in the corresponding
three-dimensional parameter spaces. The paper is concluded by section 4.

\section{The driven models}

\subsection{The parametric drive}

In the 1D case, the parametrically driven CGL equation takes the form which
generalizes the earlier studied parametrically driven damped nonlinear
Schr\"{o}dinger (NLS) equation (see Ref. \cite{Korobov}, in which exact SP
solutions to the latter equation were found, and their stability was
investigated, and further results in Refs. \cite{Barash}). The
generalization to the 2D case of the same type as in Eq. (\ref{CQ}), which
describes a planar waveguide (optical cavity) with the Kerr nonlinearity, is
straightforward: 
\begin{equation}
iu_{z}+\frac{1}{2}u_{xx}-(\frac{i}{2}+\beta )u_{tt}+\frac{1}{2}
|u|^{2}u=(k-i)u+\gamma u^{\ast },  \label{param}
\end{equation}
where the asterisk stands for the complex conjugation. The gain, which is
accounted for by the last term in Eq. (\ref{param}) with a real coefficient $
\gamma $ is provided by a pump wave copropagating with the $u$--field at the
double frequency. The real coefficient $k$ determines a wavenumber mismatch
between the signal and pump waves.

Dropping the filtering term and assuming $\beta <0$ (which physically
corresponds to anomalous dispersion of the electromagnetic waves in the
waveguide \cite{Agrawal}) transform Eq. (\ref{param}) into a 2D isotropic
damped parametrically driven NLS equation, which is used as a
phenomenological model of the Faraday ripples in hydrodynamics \cite{Faraday}
, in which case the evolutional variable $z$ in Eq. (\ref{param}) is
replaced by time $T$, and the former temporal variable $t$ is replaced by
the second transverse coordinate $y$. A more general isotropic equation,
which includes the diffusion term, 
\begin{equation}
iu_{z}+(\frac{1}{2}-iD)\nabla ^{2}u+\frac{1}{2}|u|^{2}u=(k-i)u+\gamma
u^{\ast },\,D>0,  \label{isotropic}
\end{equation}
is also used as a model to describe localized 2D states, the so-called
oscillons, observed in experiments with the Faraday resonance \cite
{Umbanhowar}. Getting back to the damped isotropic NLS equation with the
parametric drive, we note that this equation (with the evolutional variable $
z$) finds applications to the description of spatial patterns in 3D driven
optical cavities \cite{cavity-parametric}. In a recent preprint \cite
{preprint}, it has been shown that axisymmetric solitons may be stable in
the latter model, despite the presence of the collapse in its undamped
undriven version (which is the NLS equation proper).

In the present work, the objective is to find stable SPs in the strongly
anisotropic full CGL model (\ref{param}), which will have the meaning of 
{\em spatiotemporal} optical solitons \cite{Frank} propagating in the planar
waveguide. It seems obvious that pulses may only exist if the gain parameter
exceeds the loss coefficient, i.e., $\gamma >1$. On the other hand, an
obvious necessary condition for the stability of pulses is the stability of
the zero background against perturbations of the form 
\begin{equation}
u=a\exp \left( \sigma z+iqx-i\omega t\right) +b\exp \left( \sigma ^{\ast
}z-iqx+i\omega t\right)  \label{perturbation}
\end{equation}
with infinitesimal amplitudes $a$ and $b$, where $q$ and $\omega $ are
arbitrary real wavenumber and frequency of the perturbation. A dispersion
relation which determines the (complex) instability growth rate $\sigma $ as
a function of $q$ and $\omega $ can be easily found from the linearized
equation (\ref{param}): 
\begin{equation}
\left( \frac{1}{2}q^{2}-\beta \omega ^{2}+k\right) ^{2}+\left[ \left( {\rm 
Re\,}\sigma +1+\frac{1}{2}\omega ^{2}\right) +i\,{\rm Im\,}\sigma \right]
^{2}=\gamma ^{2},  \label{sigma}
\end{equation}
which gives rise to two different branches of the dispersion relation, viz., 
${\rm Re\,}\sigma =-\left( 1+\frac{1}{2}\omega ^{2}\right) $ with ${\rm Im\,}
\sigma \neq 0$, and the one with ${\rm Im\,}\sigma =0$ and $\sigma $
determined by the equation 
\begin{equation}
\left( \sigma +1+\frac{1}{2}\omega ^{2}\right) ^{2}=\gamma ^{2}-\left( \frac{
1}{2}q^{2}-\beta \omega ^{2}+k\right) ^{2}.  \label{dispersion}
\end{equation}
Obviously, the former branch always satisfies the stability condition, ${\rm 
Re\,}\sigma (q,\omega )\leq 0$. Straightforward analysis of Eq. (\ref
{dispersion}), with regard to the above assumption $\gamma >1$, shows that,
if $\omega =0$, the stability condition for the latter branch amounts to the
well-known inequalities \cite{Korobov,Barash} $\gamma ^{2}\leq 1+k^{2}$, $
k>0 $.

The consideration of Eq. (\ref{dispersion}) with $\omega \neq 0$
demonstrates that, in this case, the most dangerous perturbations are those
with $q=0$. If $\beta \leq 0$, they do not generate any additional
instability, and if $\beta $ is positive, a new stability condition appears, 
$\gamma ^{2}\leq (k+2\beta )^{2}/(1+4\beta ^{2})$; it is easy to check that
the above-mentioned necessary condition $\gamma ^{2}\leq 1+k^{2}$ is a
straightforward corollary of this inequality. Thus, the eventual system of
the conditions providing for the stability of the zero solution to Eq. (\ref
{param}) (provided that the system is infinitely extended in both $x$-- and $
t$--directions) is $k\geq 0$ and 
\begin{equation}
\begin{array}{ll}
\gamma ^{2} \leq 1+k^{2}, & {\rm if}\;\;\beta \leq 0; \\ 
\gamma ^{2} \leq (k+2\beta )^{2}/(1+4\beta ^{2}), & {\rm if} \;\;\beta >0.
\end{array}
\label{zero_stability}
\end{equation}

It is also relevant to consider nonzero spatially uniform solutions to Eq. (
\ref{param}), which have the form 
\begin{equation}
u=A_{0}\exp \left( i\phi \right) ,\,\phi =-\frac{1}{2}\sin ^{-1}(1/\gamma
),\,A_{0}=\sqrt{2\left[ k+\gamma \cos (2\phi )\right] }.  \label{uniform}
\end{equation}
This uniform solution exists for $\gamma >1$. The consideration of the
modulational stability of the solution against small perturbations with the
wavenumber $q$ and instability growth rate $\sigma $ (cf. Eq. (\ref
{perturbation})) yields a dispersion relation 
\[
\left( \frac{1}{2}q^{2}-\beta \omega ^{2}+k-A_{0}^{2}\right) ^{2}+\left[
\left( {\rm Re\,}\sigma +1+\frac{1}{2}\omega ^{2}\right) +i\,{\rm Im\,}
\sigma \right] ^{2}=\gamma ^{2}-\gamma A_{0}^{2}\cos (2\phi )+A_{0}^{4}/4. 
\]
As well as Eq. (\ref{sigma}), this equation has a solution branch with ${\rm 
Im\,}\sigma =0$, then 
\begin{equation}
\sigma =-1-\omega ^{2}/2\pm \sqrt{\gamma ^{2}-\gamma A_{0}^{2}\cos (2\phi
)+A_{0}^{4}/4-(q^{2}/2-\beta \omega ^{2}+k-A_{0}^{2})^{2}}.
\label{growthrate}
\end{equation}
The maximum value of this expression, which is attained at $\omega =0$ and $
q=\sqrt{2(k-A_{0}^{2})}$, is always positive, $\sigma _{\max }=-1+\sqrt{
1+k^{2}}$. Therefore, the uniform state (\ref{uniform}) is always unstable
against modulations in the $x$--direction, the same way as in the 1D damped
parametrically driven NLS equation \cite{Korobov,Barash}.

\subsection{The directly forced model}

The light signal in a lossy waveguide can also be directly (rather than
parametrically) supported by a pump wave launched at the same (rather than
double) frequency. For spatial patterns in 3D lossy optical cavities (a
typical example is a domain wall \cite{Maxi}), this gives rise to the
well-known isotropic directly forced 2D CGL equation, with the propagation
coordinate $z$ playing the role of the evolutional variable, see a review 
\cite{review} (a similar model describes a cavity with a trapped
Bose-Einstein condensate in the presence of source and sink of atoms \cite
{BEC}). The optical patterns generated this way may find an application in
the design of optical memory \cite{Firth}, although fully localized pulses
may be subject to collapse under the action of the self-focusing Kerr
nonlinearity \cite{Lord}.

The governing equation for the spatiotemporal optical signal in the directly
forced nonlinear waveguide differs from Eq. (\ref{param}) by the driving
term: 
\begin{equation}
iu_{z}+\frac{1}{2}u_{xx}-(\frac{i}{2}+\beta )u_{tt}+\frac{1}{2}
|u|^{2}u=(k-i)u-f,  \label{direct}
\end{equation}
where $f$ is the amplitude of the forcing field. Obviously, this equation
can only produce localized patterns on top of a constant background $u\equiv
u_{0}$, whose amplitude is determined by a cubic equation following from Eq.
(\ref{direct}), $(1/2)\left| u_{0}\right| ^{2}u_{0}+\left( i-k\right)
u_{0}+f=0$. The background can be subject to modulational instability \cite
{Agrawal}; however, if the amplitude $u_{0}$ is small enough, the
instability will be so weak that it may be disregarded for finite values of
the propagation distance that are relevant to the experiment.

\section{Numerical results}

\subsection{Two-dimensional pulses in the parametrically driven model}

Equations (\ref{param}) and (\ref{direct}) were solved numerically by means
of a pseudospectral code with $256\times 256$ modes and periodic boundary
conditions in $x$ and $t$, with a fixed size of the system in both
directions, $L_{x}=L_{t}=20$. As well as in Ref. \cite{we}, a stable 2D SP
was generated for the first time, starting with an initial Gaussian pulse
localized in both $x$ and $t$, with its center placed at the center of the
integration domain. In order to generate stability diagrams (see below), we
then varied the parameters by small steps, the initial configuration for
each simulation being the stable pulse produced by the simulation at the
previous step.

In Fig. 1 we display a typical example of a stable 2D completely localized
pulse, which was produced by the numerical solution of Eq. (\ref{param}),
following the procedure outlined above. Results of extensive simulations,
which were carried out for many different values of the parameters, can be
presented in a compact form as a stability diagram for the 2D pulses in the
three-dimensional parametric space ($\gamma ,k,\beta $), which is displayed
in Fig. 2. We stress that this parameter space comprises all the
coefficients of Eq. (\ref{param}).

The stability region proper is located under the ``roof'' in Fig. 2. To the
left of the stability border which is shown by crosses, no localized 2D
pulses can be created: in this case, an initial 2D pulse configuration
quickly decays into the stable zero state. On the other side, i.e., to the
right of the border denoted by rhombuses, localized 2D pulses were not found
either. However, the result of the evolution is different in this case: the
initial 2D pulse expands along the $t$--axis to generate a quasi-1D pulse,
which is localized in the $x$--direction, and delocalized in the $t$
--direction. Such a quasi-1D solution to Eq. (\ref{param}) can be found in
an exact form, 
\begin{equation}
u(x,t)=\sqrt{2}\kappa e^{i\phi }{\rm sech}(\kappa x),\,\phi =-\frac{1}{2}
\sin ^{-1}(1/\gamma ),\,\kappa =\sqrt{2\left[ k+\gamma \cos (2\phi )\right]}
\,.  \label{Q1D}
\end{equation}

In terms of the underlying optical models, the solution (\ref{Q1D})
corresponds to a {\it spatial} soliton. A typical example of such a soliton
is shown in Fig. 3. Straightforward comparison shows that the numerically
found quasi-1D pulse completely coincides with the analytical solution (\ref
{Q1D}). 
Simulations demonstrate that this spatial soliton is stable everywhere in
the stability region of the 2D pulses shown in Fig. 2, i.e., the model is 
{\em bistable} in this region. Beyond the right stability border of the 2D
pulses, shown by rhombuses in Fig. 2, i.e., in the region where only the
spatial solitons are found, they remain stable.

For comparison with these results, we have also performed direct simulations
of the isotropic equation (\ref{isotropic}). Stable 2D localized pulses
exist in this model too, Fig. 4 displaying an example for $D=0.1$. For $
D=0.1 $ and $k=1.3$, the stable axially symmetric 2D pulses have been found
in the interval $1.20<\gamma <1.57$. For $\gamma <1.20$, the initial 2D
pulse decays to the zero state, while for $\gamma >1.57$ the initial 2D
pulse evolves to the spatially uniform state (\ref{uniform}). In this
connection, we note that a critical branch of the modulational-instability
growth rate for this uniform solution to Eq. (\ref{isotropic}) can be found,
as a function of the perturbation wavenumber $q$, in the form $\sigma
(q)=-1-Dq^{2}+\sqrt{1+k^{2}-(q^{2}/2+k-A_{0})^{2}}$, cf. Eq. (\ref
{growthrate}). It is easy to check that, for $D=0.1$ and $k=1.3$, $\sigma
(q) $ is negative for any $q$, hence the uniform state is stable.

\subsection{Modulational instability of quasi-one-dimensional pulses in the
parametrically driven model}

In the context of the results presented above, the spatial solitons
(quasi-1D pulses) were stable. However, they become unstable against $t$
--dependent perturbations for sufficiently large negative values of the
dispersion coefficient $\beta $ in Eq. (\ref{param}), which corresponds to
anomalous dispersion in terms of nonlinear optics (in fact, modulational
instability of spatial solitons against temporal perturbations in a system
with anomalous dispersion is a general effect which has been known for a
long time \cite{Rub}). In particular, for $k=1.3$ and $\gamma =1.3$, this
instability occurs if $\beta <-1.08$. Figure 5 displays a result of
numerical simulations in this case. The initial condition including a small
perturbation is taken as 
\begin{equation}
u(x,t,z=0)=\sqrt{2}\kappa \left[ \left( 1+0.05\cos \left( \frac{8\pi t}{L}
\right) +0.01\cos \left( \frac{2\pi t}{L}\right) \right) \cos \phi +i\sin
\phi \right] {\rm sech}\left( \kappa \left( x-\frac{L}{2}\right) \right) ,
\label{z=0}
\end{equation}
where $\kappa $ and $\phi $ are the same constants as in Eq. (\ref{Q1D}).
Figure 5(a) displays the evolution of $|u(x=L/2,t)|$ at $\beta =-2.5$. While
the pattern remains localized in the $x$--direction during the evolution, it
is obvious in Fig. 5(b) that the growth of the modulation splits the
original spatial soliton into four 2D spatiotemporal pulses, but three of
them merge together, thus two 2D pulses survive (splitting of a spatial
solitary beam into temporally localized pulses under the action of
modulational instability was earlier observed in numerical \cite
{Trillo,Frank2} and laboratory \cite{Frank2} experiments with
second-harmonic-generating optical systems). Figure 5(b) displays an
eventual established state including the two 2D pulses. This modulational
instability is a generic type of transverse instability for quasi-1D pulses
(another generic type is the zigzag instability found in Ref. \cite{we} in
the context of the 2D GL equation with the CQ nonlinearity).

Another type of quasi-1D pulse solution to Eq. (\ref{param}), which is
localized in the $t$--direction and uniform in the $x$--direction, is also
possible. Such an $x$--independent solution, $u(z,t,z)=U(z,t)$, represents a 
{\it temporal} soliton in the optical cavity, and satisfies the equation 
\begin{equation}
iU_{z}-(\frac{i}{2}+\beta )U_{tt}+\frac{1}{2}|U|^{2}U=(k-i)U+\gamma U^{\ast }
\label{U}
\end{equation}
(an exact analytical solution to this equation is not available, unless $
\beta =0$). For the same case, $k=1.3$ and $\gamma =1.3$, as considered
above, this type of the quasi-1D pulse solution exists and is stable against 
$t$--dependent perturbations if $\beta <-0.812$. Figure 6(a) displays such a
pulse at $\beta =-1,k=1.3$ and $\gamma =1.3$, with its center at the point $
t=L/2$. To study its stability against $x$--dependent perturbations, we took
an initial condition 
\[
u(x,t,z=0)=\left[ 1+0.05\cos (2\pi x/L)\right] {\rm Re\,}U(t)+i{\rm Im\,}
U(t), 
\]
where $U(t)$ is the unperturbed $z$--independent SP solution to Eq. (\ref{U}
). Modulation in the $x$-- direction grows and splits the temporal soliton
into many 2D spatiotemporal pulses, as it is evident in Fig. 6(b), which
displays a 3D plot of $|u(x,t,z)|$ at $z=200$. Such a temporal soliton,
extended along the $x$--direction, seems to be always unstable against the $
x $--dependent perturbations, breaking up into a large number of 2D pulses.

One can also look for more general quasi-1D pulses of the form $
u(x,t,z)=U(x-ct)$, which may be interpreted as either moving spatial
solitons, or oblique temporal ones, the function $U$ satisfying an equation 
\[
\left( 1-2c^{2}\beta -ic^{2}\right) U^{\prime \prime
}+|U|^{2}U=2(k-i)U+2\gamma U^{\ast }. 
\]
A detailed study of such general quasi-1D solitons is beyond the scope of
this work, but one may assume that there is a critical value of the velocity 
$c$ separating completely unstable solutions and those which may be stable.

We have also simulated the development of the modulational instability of
the spatially uniform state (\ref{uniform}) in the parametrically driven
model (\ref{param}) (the presence of this instability was demonstrated
analytically in the previous section). Formation of many 2D pulses may be
expected, at sufficiently small $\beta $, as a result of the instability
development. As an initial condition, we took the uniform state (\ref
{uniform}) randomly perturbed by an initial small disturbance. Figure 7
displays a snapshot of the pattern obtained at  $z=600$ for $\beta
=-2.5,\,k=1.3$ and $\gamma =1.3$. At first, many 2D pulses are created;
however, they subsequently merge, so that only a few 2D pulses eventually
survive, which is obvious in Fig. 7.

\subsection{The directly forced model}

Numerical simulations of the directly driven CGL equation (\ref{direct})
also readily produce stable 2D pulses (built on top of the small-amplitude
background, see above). A typical example of such a pulse is shown in Fig.
8. As well as in the parametrically driven model, in this case extensive
simulations of Eq. (\ref{direct})\ have made it possible to identify a
stability region for 2D SPs in the full three-dimensional parameter space of
the model, $\left( f,k,\beta \right) $. This region is located under the
``roof'' in the parameter space, see Fig. 9. Note that the parameter space
in Fig. 9, as well as that in Fig. 2, incorporates all the parameters of the
model.

The two boundaries of the stability region in Fig. 9 have the meaning
similar to that of the stability boundaries shown in Fig. 2 for the
parametrically driven model: to the left of the boundary shown by crosses,
any initial pulse decays to zero, and to the right of the second boundary,
shown by rhombuses, the fully localized 2D pulse evolves to a quasi-1D pulse
oriented along the $t$--axis (the spatial soliton, in terms of the
underlying optical models). As well as in the parametrically driven model,
in the directly forced one the spatial solitons coexist as stable patterns
with the 2D pulses (hence, the model is bistable in that region), and they
remain stable across the right border of the stability region of the 2D
solitons, i.e., across the surface shown by rhombuses in Fig. 9.

\section{Conclusion}

In this work, we have introduced two different two-dimensional models of the
Ginzburg-Landau type with the Kerr (cubic) nonlinearity. The corresponding
equations are driven either parametrically or directly by an external field.
The models describe spatio-temporal dynamics in nonlinear dispersive lossy
optical cavities, where the loss also includes a filtering term. Both models
are strongly anisotropic, featuring only diffraction in one (spatial)
direction, and a combination of effective diffusion and dispersion in the
other (temporal) direction. By means of direct simulations, we have found
stable two-dimensional solitary pulses in both models (in the directly
driven one, the pulses are built on top of a nonzero flat background). These
two-dimensional solitary pulses correspond to spatio-temporal solitons
(``light bullets'') supported by the driving field in the planar nonlinear
waveguide.

The main results were presented in a compact form as stability regions in
the full three-dimensional parameter space of either model. The stability
region is bounded by two surfaces. Beyond one of them, any initial pulse
decays to zero, while stable quasi-one-dimensional pulses, extending along
the temporal direction (i.e., spatial solitons, in terms of the optical
cavity), were found beyond the other boundary. The spatial solitons are also
stable inside the stability regions of the two-dimensional pulses, so that
both models demonstrate bistability in these regions. Unlike the spatial
solitons, their quasi-one-dimensional counterparts in the form of temporal
solitons are always modulationally unstable. It was demonstrated that the
instability splits temporal solitons into an array of spatiotemporal
two-dimensional pulses, which, in turn, coalesce into two final pulses.
Similarly, the instability of the uniform nonzero state in the
parametrically driven model leads to formation of many two-dimensional
pulses, which subsequently merge into few two-dimensional solitons.

\section*{Figure Captions}

Fig. 1. An example of the stable stationary two-dimensional solitary pulse,
found as a solution of the parametrically driven complex Ginzburg-Landau
equation (\ref{param}). The field $\left| u\right| $ is shown vs. the
temporal and transverse spatial coordinates $t$ and $x$. Values of the
parameters are $k=1.3,\beta =-0.3,\gamma =1.30$.\newline
\newline
Fig. 2. The region of stable two-dimensional solitary pulses in the full
parameter space of Eq. (\ref{param}). The stability region is located under
the ``roof''. To the left of stability region, only zero solution is
possible, and to the right of it, the 2D pulse evolves to a stable
quasi-one-dimensional pattern, see an example in Fig. 3. \newline
\newline
Fig. 3. An example of a stable quasi-one-dimensional pulse found in the
parametrically driven complex Ginzburg-Landau equation (\ref{param}). Values
of the parameters are $k=1.3,\beta =-0.3,\gamma =1.36$ (a point
corresponding to these values is located slightly to the right of the
stability region shown in Fig. 2). 
\newline
\newline
Fig. 4 An example of a stable axisymmetric two-dimensional solitary pulse
found as a solution of the parametrically driven isotropic complex
Ginzburg-Landau equation (\ref{isotropic}). The field $\left| u\right| $ is
shown vs. the temporal and transverse spatial coordinates. Values of the
parameters are $k=1.3,D=0.1$ and $\gamma =1.30$. \newline
\newline
Fig. 5 (a) Evolution of $|u(L/2,t,z)|$ for the parametrically driven complex
Ginzburg-Landau equation (\ref{param}). The initial state is a 1D localized
pattern (\ref{z=0}) (spatial soliton) including a small modulational
perturbation, values of the parameters being $k=1.3,\beta =-2.5,\gamma =1.3$
. (b) The established stationary pattern including two 2D pulses. \newline
\newline
Fig. 6 (a) The quasi-one-dimensional pulse, which is localized in the $t$
--direction around $t=L/2$ (temporal soliton), at $\beta =-1,k=1.3$ and $
\gamma =1.3$, in the parametrically driven complex Ginzburg-Landau equation (
\ref{param}). (b) The result of the instability development of this soliton,
shown as a plot of $|u(x,t)|$ at $z=200$. \newline
\newline
Fig. 7 Patterns produced by the modulational instability of the uniform
state in the parametrically driven complex Ginzburg-Landau equation (\ref
{param}) at $z=600$. Values of parameter are $\beta =-2.5,k=1.3$ and $\gamma
=1.3$.\newline
\newline
Fig. 8. An example of the stable stationary two-dimensional solitary pulse,
found on top of the nonzero flat background, in the directly forced complex
Ginzburg-Landau equation (\ref{direct}). The field $\left| u\right| $ is
shown vs. the temporal and transverse spatial coordinates $t$ and $x$.
Values of the parameters are $k=3,\beta =-0.3,f=3.1$. \newline
\newline
Fig. 9. The region of stable two-dimensional solitary pulses (built on top
of the nonzero flat background) in the full parameter space of the directly
forced complex Ginzburg-Landau equation (\ref{direct}). The stability region
is located under the ``roof''. To the left of stability region, the 2D pulse
decays to the flat-background solution, and to the right of the stability
region, the 2D pulse evolves to a stable quasi-one-dimensional pattern
(spatial soliton).

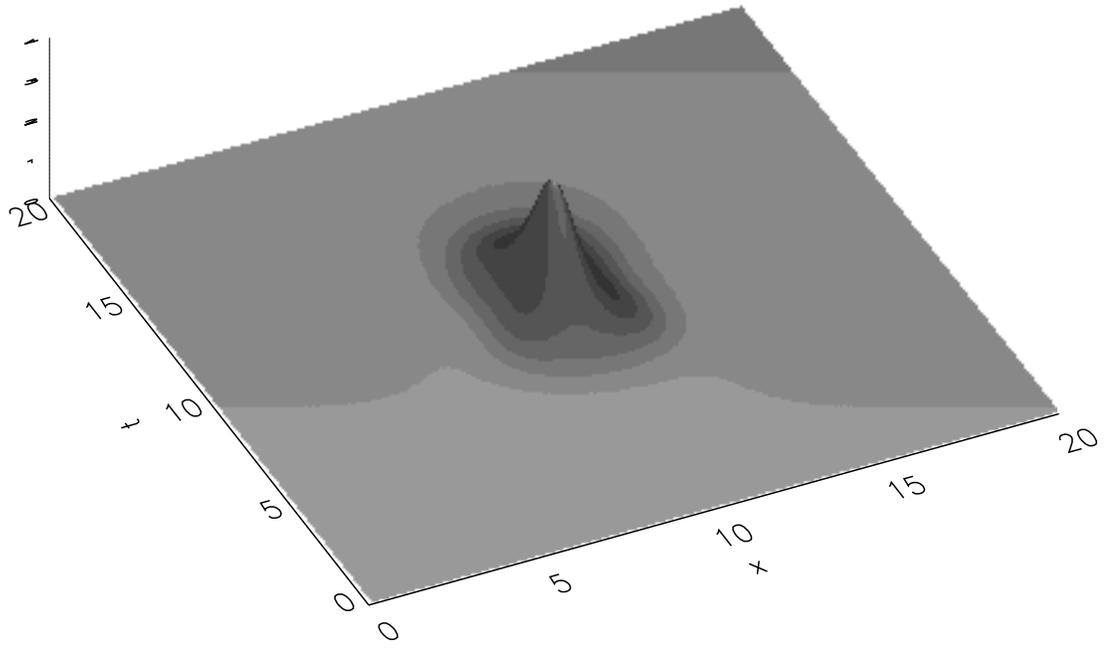
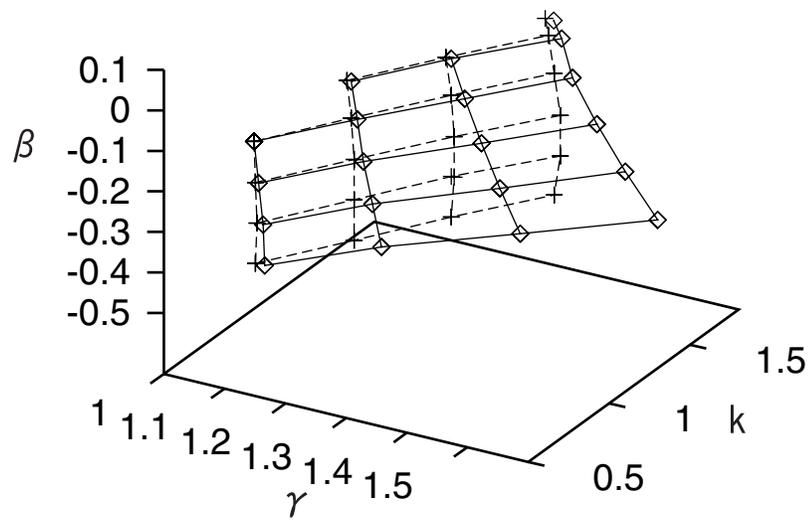
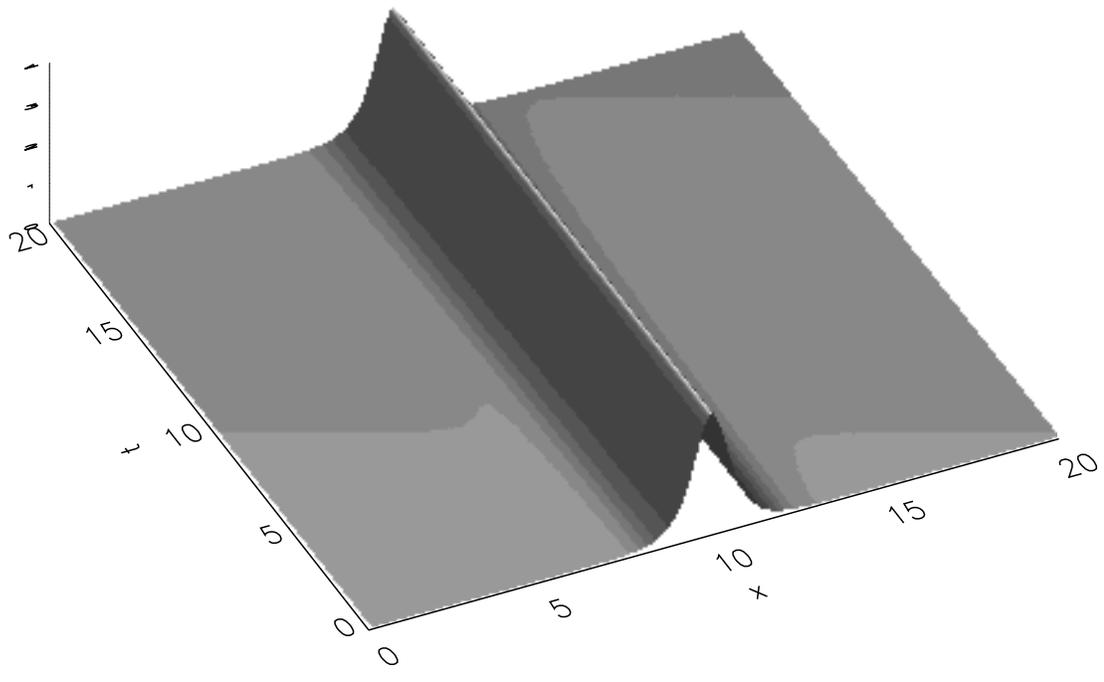
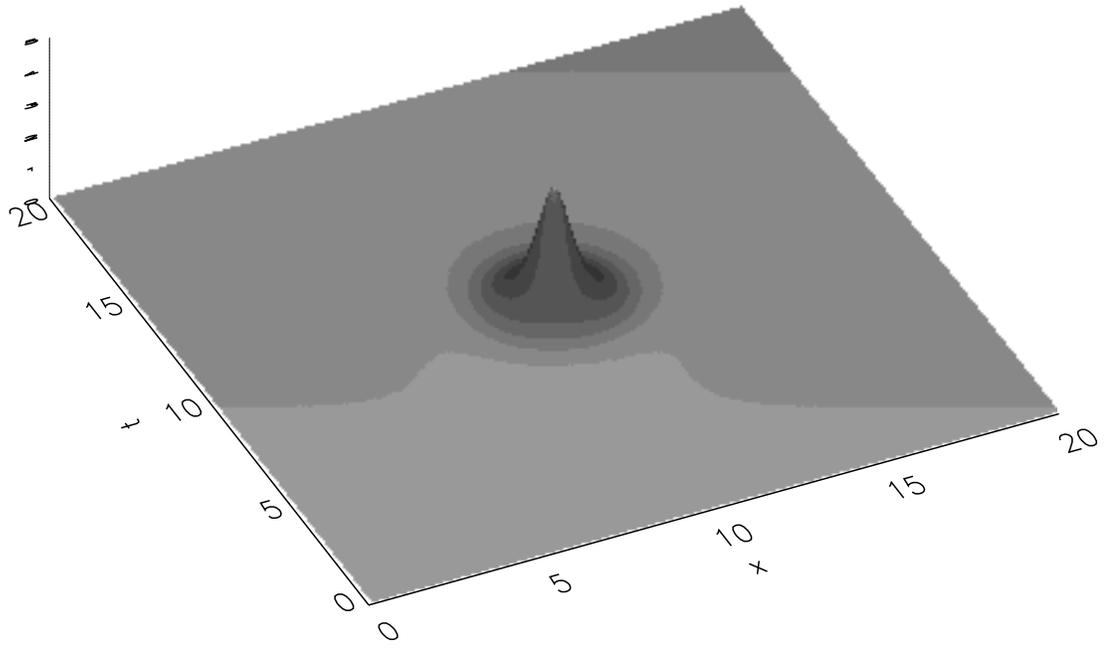
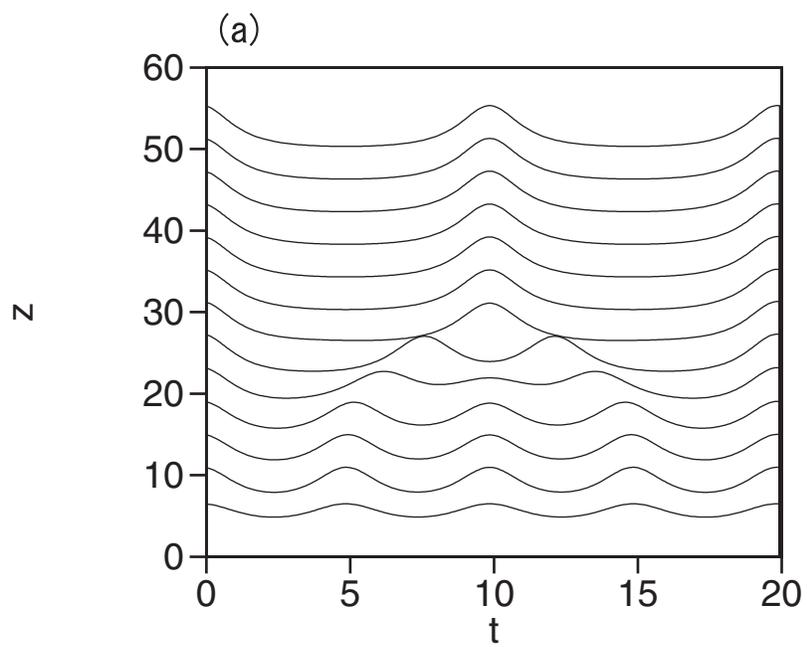
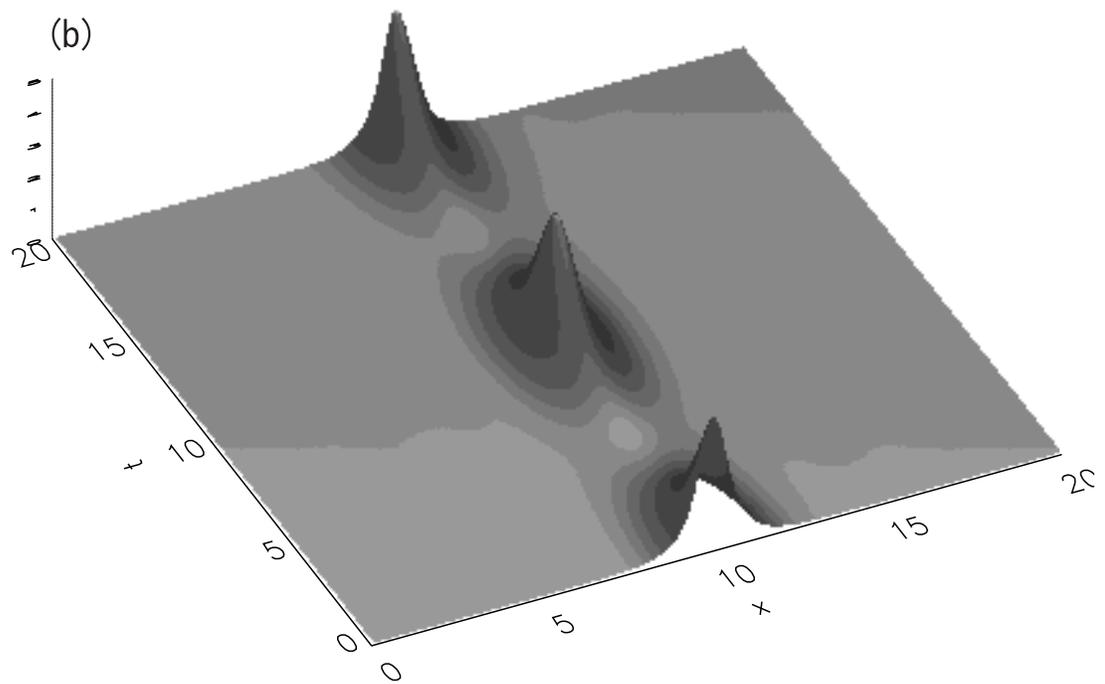
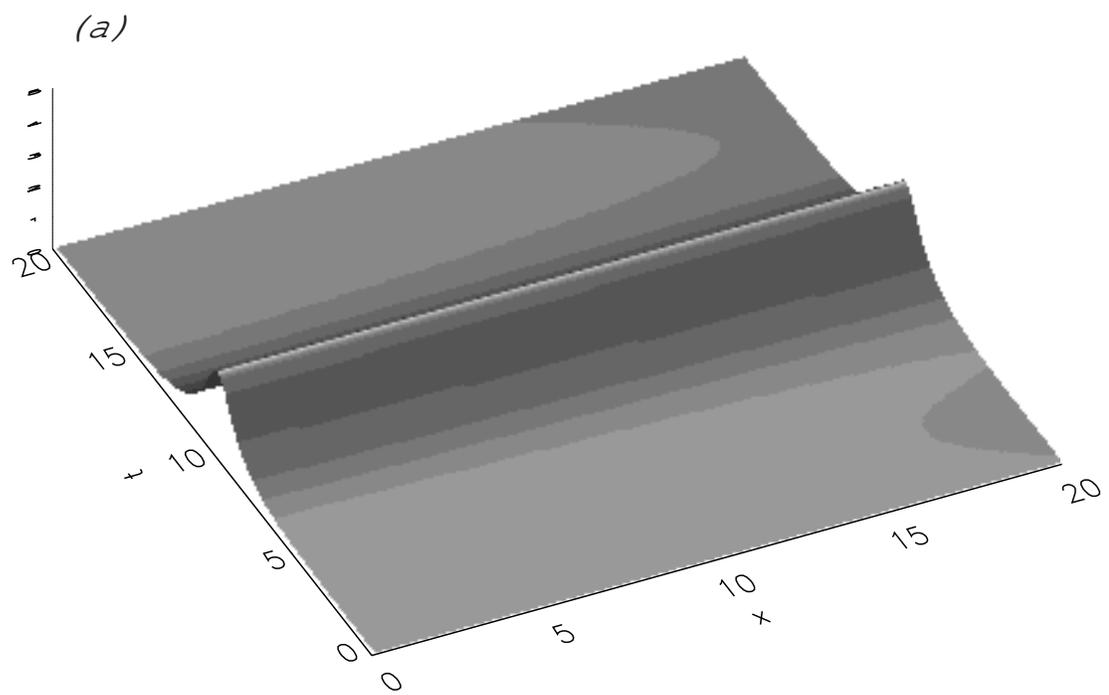
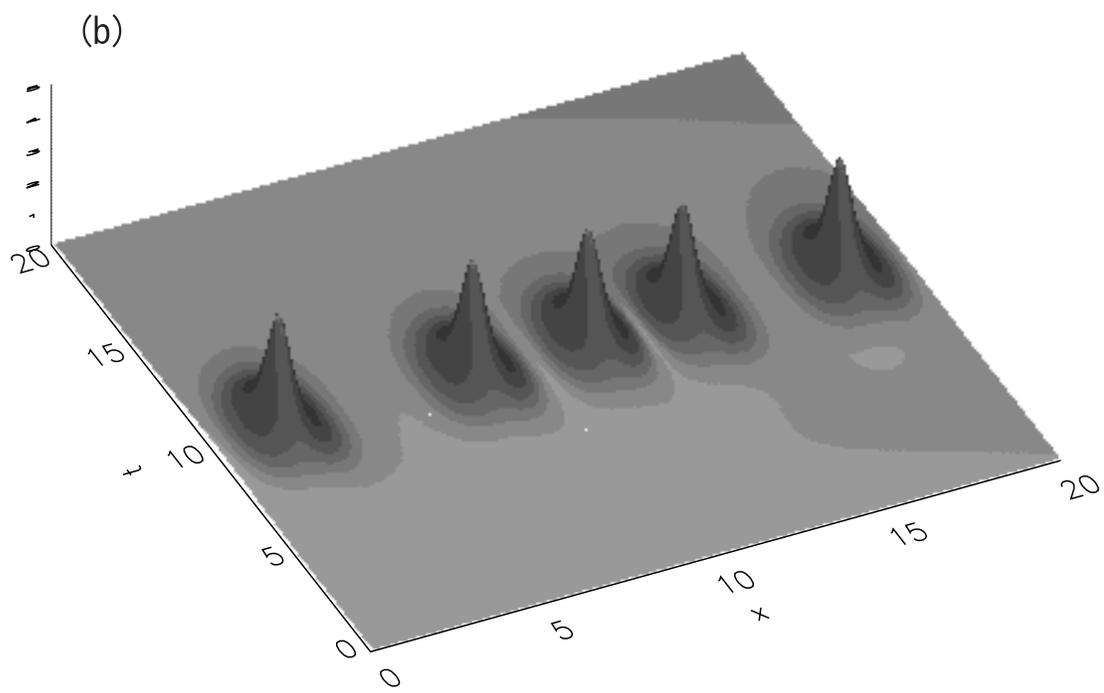
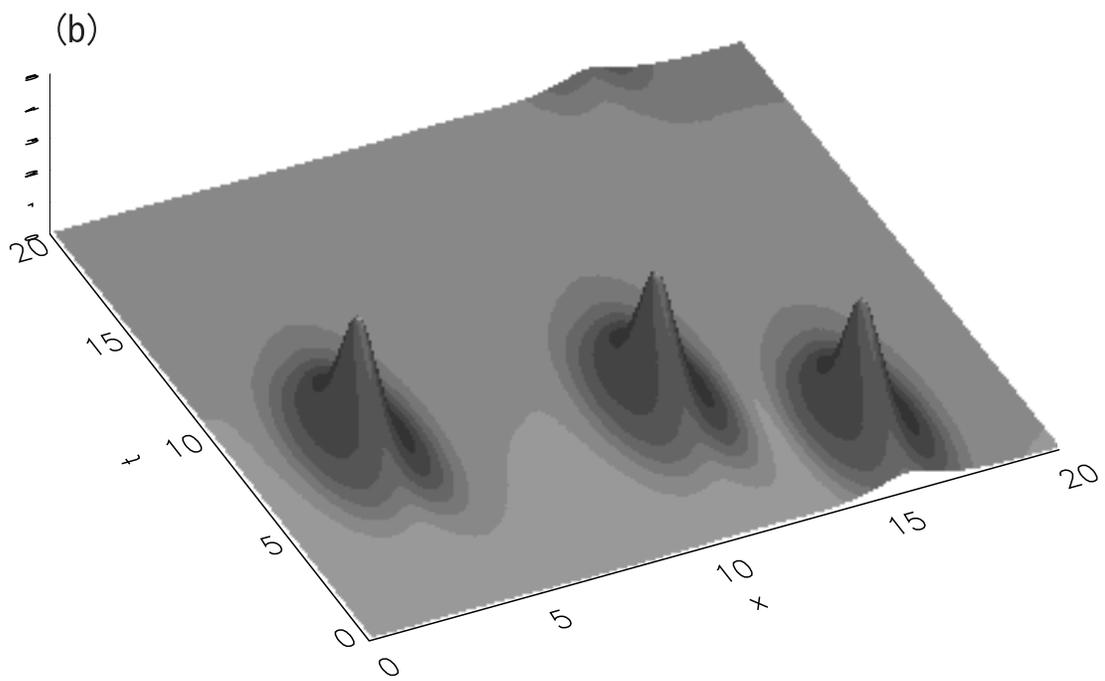
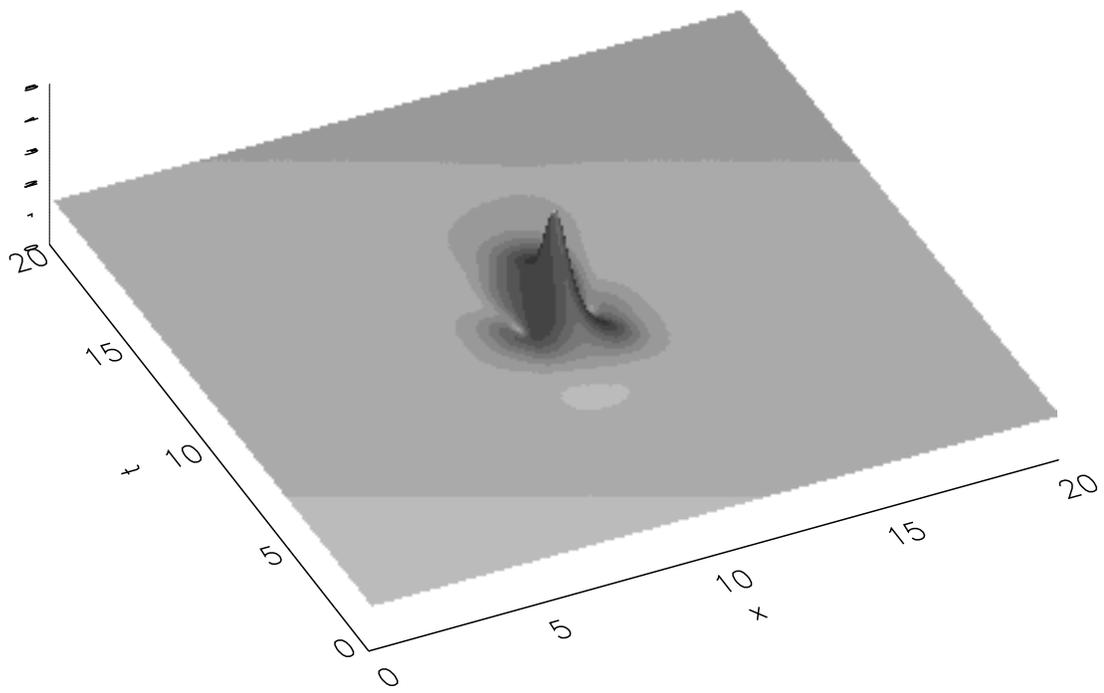
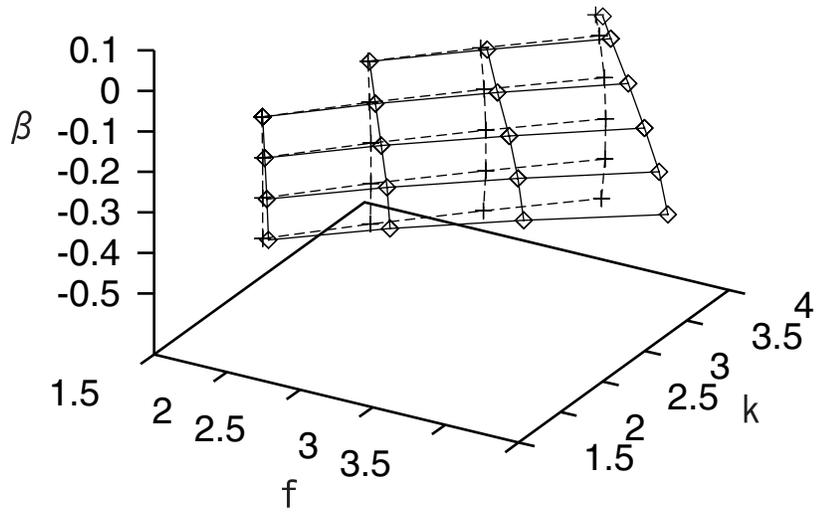
\end{document}